\documentclass[linenumbers,twocolumn]{aastex63}
\nolinenumbers

\newcommand{\add}[1]{{#1}}
\shorttitle{Scale and Time Dependence of Alfvénicity}
\shortauthors{Thepthong et al.}
\graphicspath{{./}{figures/}}

\begin{document}

\title{Scale and Time Dependence of Alfvénicity in the Solar Wind as Observed by {\it Parker Solar Probe}}


\author[0000-0001-9597-1448]{Panisara Thepthong}
\affiliation{Department of Physics, Faculty of Science, Kasetsart University, Bangkok 10900, Thailand}

\author[0000-0002-6609-1422]{Peera Pongkitiwanichakul}
\affiliation{Department of Physics, Faculty of Science, Kasetsart University, Bangkok 10900, Thailand}

\author[0000-0003-3414-9666]{David Ruffolo}
\affiliation{Department of Physics, Faculty of Science, Mahidol University, Bangkok 10400, Thailand}

\author[0000-0003-0937-2655]{Rungployphan Kieokaew}
\affiliation{Institut de Recherche en Astrophysique et Planétologie, CNRS, UPS, CNES, 9 Ave. du Colonel Roche, 31028 Toulouse, France}

\author[0000-0002-6962-0959]{Riddhi Bandyopadhyay}
\affiliation{Department of Astrophysical Sciences, Princeton University, Princeton, NJ 08544, USA}

\author[0000-0001-7224-6024]{William H. Matthaeus}
\affiliation{Bartol Research Institute and Department of Physics and Astronomy, University of Delaware, Newark, DE 19716, USA}

\author[0000-0003-0602-8381]{Tulasi N. Parashar}
\affiliation{School of Chemical and Physical Sciences, Victoria University of Wellington, Wellington 6012,
New Zealand}

\begin{abstract}
\nolinenumbers
Alfv\'enicity is a well-known property, common in the solar wind, characterized by a high correlation between magnetic and velocity fluctuations. 
Data from the Parker Solar Probe (PSP) enable the study of this property closer to the Sun than ever before, as well as  in sub-Alfv\'enic solar wind.  
We consider scale-dependent measures of Alfv\'enicity based on second-order functions of the magnetic and velocity increments as a function of time lag, including the normalized cross-helicity $\sigma_c$ and residual energy $\sigma_r$.
Scale-dependent Alfv\'enicity is strongest for lags near the correlation scale and increases when moving closer to the Sun.
We find that $\sigma_r$ typically remains close to the maximally negative value compatible with $\sigma_c$.
We did not observe significant changes in measures of Alfv\'enicity between sub-Alfv\'enic and super-Alfv\'enic wind.
During most times, the solar wind was highly Alfv\'enic; 
however, lower Alfv\'enicity was observed when PSP approached the heliospheric current sheet or other magnetic structures with sudden changes in the radial magnetic field, non-unidirectional strahl electron pitch angle distributions, and strong electron density contrasts. These results are consistent with a picture in which Alfv\'enic fluctuations generated near the photosphere
transport outward forming highly Alfv\'enic states in the young solar wind and subsequent interactions with
large scale structures and gradients leads to weaker Alfv\'enicity, as commonly observed at larger heliocentric distances.

\end{abstract}

 

\accepted{for publication in the {\it Astrophysical Journal}}
 
\section{Introduction} \label{sec:intro}

The solar wind is strongly turbulent and often exhibits an obvious correlation between large fluctuations in velocity and magnetic field over a wide frequency range \citep[e.g.,][]{Belcher_Davis_1971,KasperEA19}. This property is often referred to as ``Alfvénicity'' in view of the resemblance of these fluctuations to large amplitude Alfvén waves \citep{Barnes79}. Often Alfvénic wind also displays a low level of density fluctuations and a nearly constant magnitude of the magnetic field vector \citep{Barnes81}. The domains of nearly constant magnetic fields describe magnetic pressure balance, and these domains are separated by significant changes in magnetic field magnitude \citep{Ruffolo2021}.  The degree of Alfvénicity in the solar wind can vary depending on location and its source \citep{DAmicisEA21}. Alfv\'enicity tends to be higher at higher heliolatitude \citep{McComas_2000}, in the higher-speed solar wind at low latitude \citep{Bruno_2003}, and closer to the Sun \citep{Chen2020}. 

These Alfv\'enic properties are features of large-amplitude and low-frequency Alfvén waves, with velocity fluctuations {\bf v} and magnetic field fluctuations {\bf b} connected through the Walén relation \citep{Walen_1944} 
\begin{equation}
    {\bf v}=\pm \frac{\bf b}{\sqrt{\mu_0\rho}},
    \label{eq:walen}
\end{equation}
where $\mu_0$ is the vacuum permeability, and $\rho$ is plasma density.
Even with a large amplitude, these incompressive waves are solutions of the compressible magnetohydrodynamic (MHD) equations, provided that the magnitude of the total (background plus fluctuating) magnetic field is constant 
\citep{Goldstein1974, Barnes79}.  Based on this reasoning it is commonly assumed that an ensemble of large-amplitude Alfv\'en wave packets of this type comprise much of the turbulence observed in the solar wind.
Moreover, the strong Alfvénicity observed in the solar wind is usually associated with large amplitude wave packets that are also polarized in the sense of outward propagation \citep{Belcher_Davis_1971}. 


Parker Solar Probe (PSP) \citep{Fox_2016} is a spacecraft launched in 2018 to study the Sun using a combination of in-situ and remote sensing measurements to study phenomena in the solar wind as well as energetic particles from solar storms and other sources. PSP enables us to explore the Sun closer than any previous spacecraft, providing the opportunity to study the features of Alfvénic turbulence in its early stages, including the high correlation between velocity and magnetic field fluctuations, as well as the nearly constant magnitude of the magnetic field. 
During its 8th orbit (also called Solar Encounter 8 or E8) at a perihelion distance of around 0.074 AU (16 $R_\odot$), PSP became the first spacecraft to provide {\it in situ} observations of sub-Alfvénic solar wind \citep{KasperEA21},
where the solar wind speed is lower than the Alfvén speed, $V_A$, given by the following equation:
\begin{equation}
V_A = \frac{|\bf{B}|}{\sqrt{\mu_0m_pn_e}},
\label{eq;V_A}
\end{equation}
where $|\bf{B}|$ is the magnitude of the magnetic field, $m_p$ is the proton mass, and $n_e$ is the electron density of the plasma  
(assuming plasma neutrality and neglecting the effect of the different mass-to-charge ratios of minor ions).  Adopting the term ``Alfv\'en critical zone'' to refer to the spatial region where $V\sim V_A$ \citep{DeForestEA18,ChhiberEA22}, which has also been called the Alfv\'en critical point or surface, 
PSP now makes it possible to directly examine Alfv\'enicity near and inside this zone.
This can help us gain a better understanding of the origin and nature of solar wind turbulence and the mechanisms that accelerate the solar wind and heat the solar corona.

In this work, we analyze data from multiple orbits of the Parker Solar Probe mission down to the distance of 0.062 au from the Sun. 
We determine quantitative measures of Alfvénicity from increments of the velocity and magnetic field. 
We explain the relationship between second-order functions of vector field increments and the corresponding Fourier spectra. 
We then analyze the scale dependence of Alfvénicity and investigate the association between Alfvénicity and other parameters as a function of distance and time. 
We also examine the time periods during which PSP observed anomalously low Alfvénicity and discuss possible mechanisms that can reduce Alfvénicity in the solar wind.

\section{Data} \label{sec:data}
We used magnetic field ($\bf{B}$) data from the flux-gate magnetometer of the PSP/FIELDS instrument suite \citep{Bale_2016} and proton partial moment ($\bf{V}$) and temperature ($T$) data from the Solar Probe ANalyzer for Ions (SPAN-i) instrument in the PSP/SWEAP instrument suite \citep{Kasper_2016}.  
The electron density ($n_e$) is derived from the quasi-thermal noise spectrum measured by the  PSP/FIELDS Radio Frequency Spectrometer \citep{Moncuquet_2020}, and the electron strahl pitch angle distribution is measured by the Solar Probe ANalyzer for Electrons (SPAN-e) of PSP/SWEAP \citep{Whittlesey_2020}. 
To enable a consistent comparison of the data, we resampled all FIELDS and SWEAP data to a cadence of 1 NYs (a ``New York second'' equal to 0.874 s), the native cadence of SWEAP, interpolating the two sets of data to the time stamp of the SWEAP data.

In this work, we analyzed public data during PSP encounter 8 from 2021 April 24 00:00 UTC to May 4 18:00 UTC, encounter 9 from 2021 August 4 00:00 UTC to August 14 23:59 UTC, and encounter 10 from 2021 November 15 00:00 UTC to November 25 23:59 UTC. During these time intervals, PSP observed several time periods of sub-Alfvénic solar wind \citep{KasperEA21}, characterized by an Alfvén Mach number $M_A$ less than 1. 
The Alfvén Mach number is calculated as the ratio of the proton radial velocity $V_R$  to the local Alfvén speed $V_A$.

\section{Analysis Techniques}

\subsection{Increments and Structure Functions \label{3.1}}

The scale-dependent measures used to quantify the Alfvénicity of the solar wind are similar to those described by 
\cite{Parashar2020}. In order to include a sensitivity to scale, the  following measures will be based on 
{\it increments} and structure functions (or other second-order functions of increments) rather than on
ordinary fluctuations that are defined in terms of departures from a mean field \citep[see, e.g.,][]{MatthaeusGoldstein82meas}. 
These measures include the normalized increment 
cross helicity, the increment Alfvén ratio, the normalized increment residual energy, and the alignment cosine between velocity and magnetic field increments. 
Following the definitions below,
for the remainder of
the paper we will drop the modifier ``increment'' from these terms, with the understanding that we are employing 
increment-based measures. 

The solar wind velocity and magnetic increments for time lag $\tau$ are described as
\begin{equation} 
    {\Delta \bf{V}} = {\bf V}(t + \tau) - {\bf V}(t)
\label{eq:dv}
\end{equation}
\begin{equation} 
    {\Delta \bf{B}} = {\bf B}(t + \tau) - {\bf B}(t)
\label{eq:db}
\end{equation}
where ${\Delta \bf{B}}$ is usually measured in Alfvén speed units with implied division by  $\sqrt{\mu_0m_pn_e}$. 

In certain circumstances, it would be useful to instead define the increments in terms of fluctuating fields
${\bf v} \equiv {\bf V} - \langle {\bf V} \rangle$
and ${\bf b} \equiv {\bf B} - \langle {\bf B} \rangle $.
Such increments would differ from those in Equations (\ref{eq:dv}) and (\ref{eq:db}) when the mean fields are time-dependent, e.g., based on running averages, a difference that becomes more pronounced at larger $\tau$.
In the present work, we use Equations (\ref{eq:dv}) and (\ref{eq:db}) directly and avoid the need for averaging.
Indeed, avoiding an averaging or detrending procedure has been considered a substantial advantage of using structure functions rather than Fourier spectra of turbulent fluctuations \citep{Lindborg99}.
In any case, at large $\tau$, the spacecraft locations at times $t$ and $t+\tau$ are often in distinct plasma streams with different physical features, so Taylor's frozen-in hypothesis \citep{Taylor38} is a poor approximation and this technique, or indeed any single-spacecraft technique, 
is no longer providing information about the spatial properties of local turbulence; rather it informs us about temporal decorrelation as influenced by such stream crossings.

As a first example, we define the 
increment cross helicity as 
\begin{equation} 
    H_c = {\langle {\Delta \bf{V}} \cdot \Delta \bf{B} \rangle}.
\label{Eq:Hc}
\end{equation}
It is straightforward to see that,
so defined, this quantity behaves qualitatively as the second-order structure functions 
that are familiar in hydrodynamics  \citep[e.g.,][]{MoninYaglom,Frisch95}. 
Note that the bracket $\langle \dots \rangle$ introduced in equation ($\ref{Eq:Hc}$) represents a suitable time average appropriate to local solar wind conditions, and in this work we average for each value of $t$ used in Equations (\ref{eq:dv}) and (\ref{eq:db}), with a cadence of 1 NYs, over non-overlapping time intervals, usually of 10 minutes.
This quantity measures the correlation between velocity and magnetic increments and is therefore relevant to assessing Alfv\'enicity (see Equation \ref{eq:walen}).
As will be explained later, the sign of $\Delta \bf{V} \cdot \Delta \bf{B}$ typically reverses with the sign of $B_r$, i.e., with magnetic sector crossings.
To avoid including both signs  within the averaging of Equation (\ref{Eq:Hc}), in which case the result would depend strongly on the durations of magnetic sectors within the averaging period, we sometimes use a rectified cross helicity defined by
\begin{equation} 
    \tilde H_c = {\langle |{\Delta \bf{V}} \cdot \Delta \bf{B}| \rangle}.
\label{Eq:rect-Hc}
\end{equation}

Similar to $H_c$, one may define 
a second-order structure function separately for the
magnetic field vector increments, as
\begin{equation} 
    S_b = {\langle {\Delta \bf{B}} \cdot \Delta \bf{B} \rangle}
\label{eq:Sb}
\end{equation}
and for the 
velocity field vector increments, as
\begin{equation} 
    S_v = {\langle {\Delta \bf{V}} \cdot \Delta \bf{V} \rangle}.
\label{eq:Sv}
\end{equation}
Note that all of these second-order quantities depend on the time 
lag $\tau$ as an argument, which is suppressed here. 

These quantities begin at zero lag at a value of zero, and reach an asymptotic value at large lag to twice the estimated value of 
the total $\langle {\bf v} \cdot {\bf b}\rangle$, 
$\langle {\bf b^2} \rangle$, or 
$\langle {\bf v^2}\rangle$
defined in terms of the fluctuating fields
${\bf v} \equiv {\bf V} - \langle {\bf V} \rangle$
and ${\bf b} \equiv {\bf B} - \langle {\bf B} \rangle $.
A relevant physical interpretation is that each 
function represents the contribution to its respective asymptotic value (energy or cross helicity) due to all fluctuations at time scales less than the lag $\tau$ \citep[e.g.,][]{Davidson15}. 

This of course corresponds to the behavior of standard second-order structure functions in homogeneous hydrodynamic turbulence. 
We also note that these functions are related to
rugged invariants of incompressible magnetohydrodynamic turbulence 
\citep[see, e.g.,][]{MatthaeusGoldstein82meas}.

\begin{figure*}
  \includegraphics[width=\linewidth]{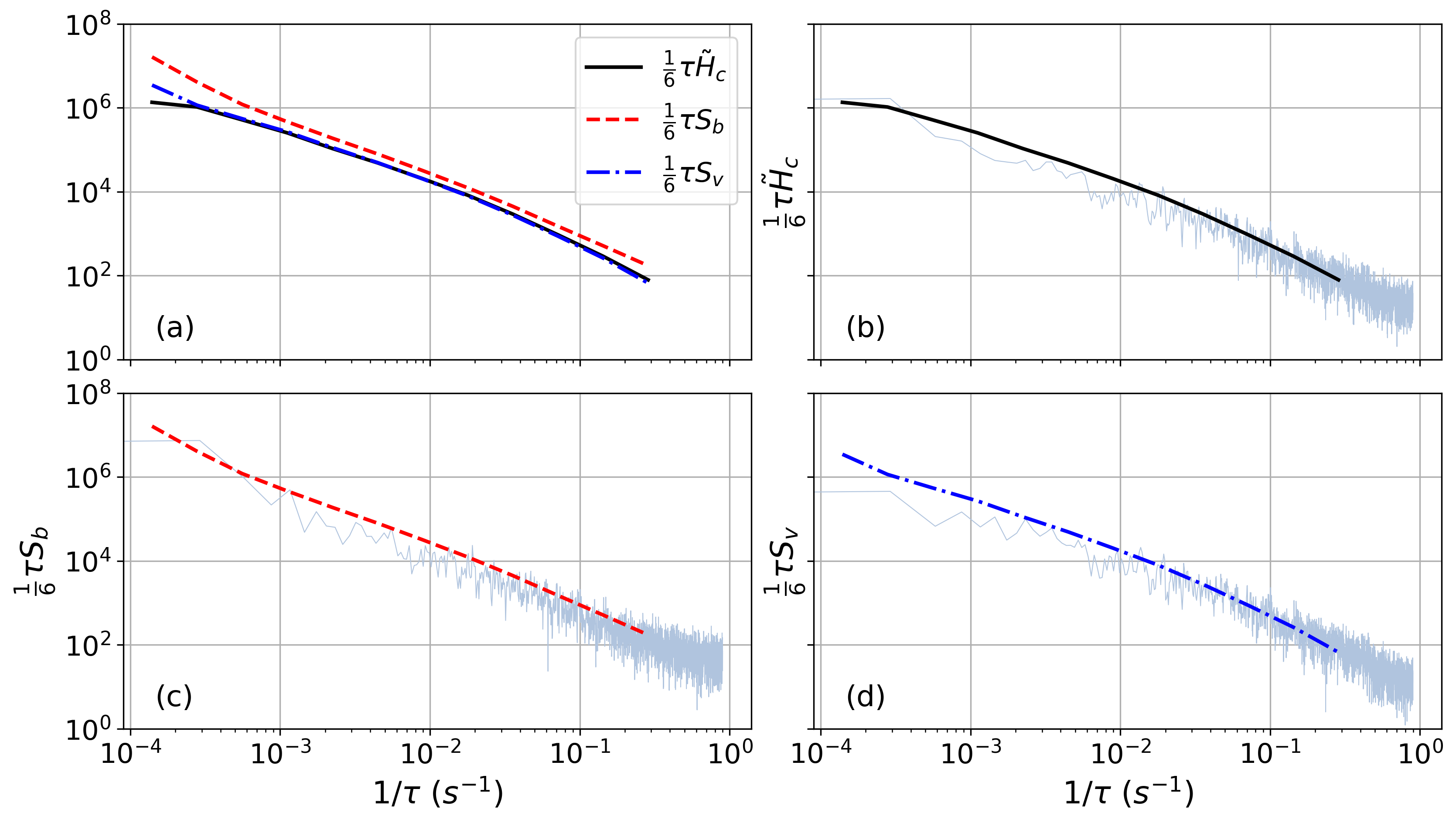}
  \caption{Equivalent spectra computed from increments at time lag $\tau$ of solar wind velocity and magnetic field using PSP data from a sample interval from 10:00 to 16:00 UTC on 2021 April 29 at a distance of 0.075 au from the Sun. 
  Panel (a) shows equivalent spectra of rectified cross helicity (black solid curve),  magnetic field (red dashed curve), and velocity (blue dash-dotted curve). Panels (b)-(d) display equivalent spectra for each individual second-order function, along with the corresponding Fourier spectrum (light blue curve).
  The equivalent spectra provide information similar to Fourier spectra, with less scatter.
  The observation that $S_b>S_v$ at each time scale $\tau$, indicating dominance of magnetic energy over kinetic energy, is a well-known feature of solar wind turbulence, which can be equivalently expressed in terms of a low Alfv\'en ratio, $r_A<1$, and negative residual energy, $\sigma_r<0$.}
  \label{fig:Structure}
\end{figure*}

\subsection{Relations to Spectra \label{3.2}}
Some previous work \citep[e.g.,][]{ChasapisEA17,ChhiberEA2018,Parashar2018} has considered a structure function, computed from increments, in a form called an ``equivalent spectrum'' that is directly comparable to, but with less scatter than, a Fourier spectrum.  
We can explain the relationship as follows:

The Fourier spectrum \add{$\bf{P(\omega)}$} for, say, magnetic fluctuations about a mean value can be normalized to satisfy the relation
\begin{equation} 
    \langle {b^2} \rangle = 2\int_{0}^{\infty} P(\omega) d\omega.
\label{eq:deltab2}
\end{equation}
In terms of the structure function $S_b(\tau)$, as $\tau\to\infty$ there is no correlation between ${\bf b}(t + \tau)$ and ${\bf b}(t)$, so 
\begin{equation} 
    \langle {b^2} \rangle = \frac{S_b(\infty)}{2} = 
    \int_{0}^{\infty} 
    \frac{1}{2}\frac{d\add{S_b}}{d\tau}d\tau.
\label{eq:dS}
\end{equation}
Now we identify $\omega$ with $1/\tau$, and therefore express the integral in terms of $1/\tau$:
\begin{equation} 
    \langle {b^2} \rangle = 
    2\int_{0}^{\infty} 
    \frac{1}{4}\tau^2\frac{d\add{S_b}}{d\tau}d\left(\frac{1}{\tau}\right).
\label{eq:gen}
\end{equation}
\add{The above derivation also applies if $S_b$ is replaced by $S_v$ or $\tilde H_c$; in the following we use $S$ to refer to any of these quantities.}
Comparing Equations (\ref{eq:deltab2}) and (\ref{eq:gen}), $(1/4)\tau^2 dS(\tau)/d\tau$ as a function of $1/\tau$ can be directly compared with, and be interpreted in the same way as, the Fourier spectrum.
As noted above, $S(\tau)$ expresses the cumulative contributions of fluctuations at all time scales up to $\tau$, so $dS/d\tau$ relates to the specific contribution from the time scale $\tau$.
Therefore we identify 
\begin{equation}
S_{eq}(\tau)\equiv\frac{\tau^2}{4} \frac{dS(\tau)}{d\tau}
\end{equation}
as an equivalent spectrum, based on structure functions.

Alternatively, we can define a function $\tilde S(\omega)\equiv S(1/\tau)$, giving us another expression of the equivalent spectrum:
\begin{equation}
P_{eq}(\omega)=-\frac{1}{4}\frac{d\tilde S(\omega)}{d\omega}.
\end{equation}

According to Figure \ref{fig:Structure}, our second-order functions often have a power-law form $S \propto \tau^{\alpha}$, and we find that they can be described by the Kolmogorov index $\alpha=2/3$, at least over certain ranges of $\tau$.  
Then we can approximate the equivalent spectrum as $(1/6)\tau S(\tau)$.


Figure \ref{fig:Structure} compares the approximate equivalent spectra $(1/6)\tau S(\tau)$ based on second-order 
functions of rectified cross-helicity, magnetic field, and velocity (colored, dashed traces) with the corresponding Fourier transforms (light blue).
Each equivalent spectrum is similar to but much smoother than the corresponding Fourier transform.

\subsection{Measures of Alfv\'enicity}

At this point, we introduce normalized quantities that are formed as ratios involving the above elementary second-order functions of increments, and which provide physical interpretations related to solar wind properties. See \cite{BrunoCarbone13} for other applications
of increments to solar wind studies.

\begin{figure*}
\centering
  \includegraphics[width=\linewidth]{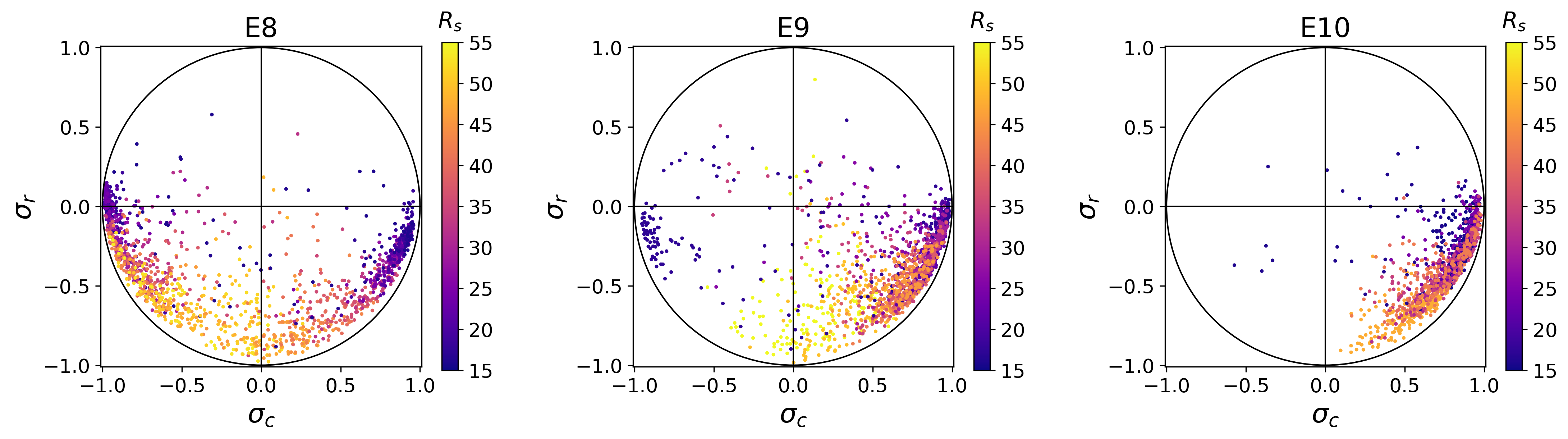}
  \caption{Normalized cross helicity as a function of normalized residual energy for PSP Solar Encounters 8, 9 and 10, using velocity and magnetic field increments over a lag of $\tau=$ 87.4 s. Each data point is based on averages over a non-overlapping 10-minute time window. The color scale represents the heliocentric distance from the Sun in solar radii ($R_s$). The black circle indicates the mathematical constraint $\sigma_c^2$ + $\sigma_r^2$ = 1. }
  \label{fig:sigmac_vs_sigmar}
\end{figure*}

The normalized 
cross helicity is defined as
\begin{equation} 
    \sigma_c = \frac{2\langle {\Delta \bf{V}} \cdot \Delta \bf{B} \rangle}{\langle {|\Delta \bf{V}|}^2 \rangle + \langle {|\Delta \bf{B}|}^2 \rangle}.
\label{eq;sigma_c}
\end{equation}
Note that $\langle {|\Delta \bf{V}|}^2 \rangle + \langle {|\Delta \bf{B}|}^2 \rangle \pm 2 \langle {\Delta \bf{V}} \cdot \Delta \mathbf{B} \rangle \geq 0$, so $-1\leq\sigma_c\leq1$.
This measure allows us to assess the cumulative degree of correlation between the velocity and magnetic field fluctuations, a characteristic of Alfv\'enic fluctuations, as a function of scale. 
To be precise, recalling that a structure function or related quantity represents the cumulative contribution of fluctuations at scales below and up to the lag $\tau$, $\sigma_c$ is the ratio of the cumulative cross helicity to the cumulative energy up to the scale $\tau$.
A value of $|\sigma_c|$close to 1 is indicative of highly Alfvénic fluctuations.

The Alfvén ratio is the ratio of the velocity field structure function to the magnetic field structure function: 
\begin{equation} 
    r_A = \frac{\langle {|\Delta \bf{V}|}^2 \rangle}{\langle {|\Delta \bf{B}|}^2 \rangle}.
\label{eq;r_A}
\end{equation}
It represents the ratio of cumulative kinetic energy density to cumulative magnetic fluctuation energy density up to the scale $\tau$. 
In a spectrum of normal unidirectionally propagating Alfvén waves, the value of $r_A$ is equal to 1; other values indicate deviation from ideal Alfv\'enicity.
    
Another measure that we use to characterize Alfvénicity in solar wind plasma is the normalized residual energy, which combines information of the Alfv\'en ratio and the fluctuation energy.
It is normally defined as the difference in kinetic and magnetic energies normalized by their sum. In keeping with the current approach based on structure functions and related quantities, it is defined here as 
\begin{equation} 
    \sigma_r = \frac{\langle {|\Delta \bf{V}|}^2 \rangle - \langle {|\Delta \bf{B}|}^2 \rangle}{\langle {|\Delta \bf{V}|}^2 \rangle + \langle {|\Delta \bf{B}|}^2 \rangle}.
\label{eq;sigma_r}
\end{equation}
For Alfvén waves, and for ideal Alfv\'enicity, there is equal energy in velocity and magnetic fluctuations and the value of $\sigma_r$ is equal to 0. However, solar wind observations usually indicate a negative value \citep{MatthaeusGoldstein82meas,BrunoEA85}.

The final measurement is the global alignment cosine angle between the magnetic and velocity fluctuations. This measure is defined as
\begin{equation} 
    \cos{\Theta} = \frac{\langle {\Delta \bf{V}} \cdot {\Delta \bf{B}} \rangle}{\sqrt{\langle {|\Delta \bf{V}|}^2 \rangle \langle {|\Delta \bf{B}|}^2 \rangle}}.
\label{eq;cos_vb}
\end{equation}
This indicates the degree of directional alignment between the velocity and magnetic field fluctuations. A value of 1 indicates that the two quantities are perfectly aligned, while a value of -1 indicates that they are perfectly anti-aligned.  
Either of these values is consistent with ideal Alfv\'enicity.

These measures of Alfv\'enicity are based on second-order functions of increments, and their contributions come from fluctuations at scales smaller than $\tau$. Power spectra in the solar wind are observed to rise with increasing $\tau$ (see Figure \ref{fig:Structure}), so the main contribution to such a second-order function comes from fluctuations at scales close to $\tau$, and our measures of Alfvénicity are generally dominated by fluctuations of scale $\sim\tau$. 

We also note that these quantities are not all independent.  
For example, 
\begin{equation} 
    \sigma_r = \frac{r_A - 1}{r_A + 1}
\label{eq;sigma_r 2}
\end{equation}
and 
\begin{equation} 
    \sigma_c = \cos{\Theta} \sqrt{1-{\sigma_r}^2},
\label{eq;sigma_c2}
\end{equation}
which implies that
\begin{equation} 
    \sigma_r^2+ \sigma_c^2 \leq 1
\label{eq:sigma-r-c-constraint}.
\end{equation}
Equivalently \citep{MatthaeusGoldstein82meas},
\begin{equation} 
    \sigma_c = 2\cos{\Theta} \frac{\sqrt{r_A}}{1+ r_A}.
\label{eq;sigma_c 3}
\end{equation}

The Elsässer increments are defined as 
\begin{equation} 
     \Delta\textbf{Z}^{\pm} = \Delta \textbf{V} \pm\Delta \textbf{B}
\label{eq:elsasser}
\end{equation}
where $\bf{Z^-}$ is the Els\"asser field that propagates parallel to the mean magnetic field and $\bf{Z^+}$ has anti-parallel propagation.
In further analysis, the scalar notation $\Delta Z^\pm$ refers to the rms value of this vector increment.

\section{Results} \label{sec:results}

In this section, we present the results of our analysis, where we utilize data processed in terms of increments to quantify Alfvénicity. 
We examine four main issues: its relation to distance from the Sun, scale-dependent Alfvénicity, time dependence of Alfvénicity near perihelia, and the occurrence of time periods of low Alfvénicity. 

\begin{figure*}
  \includegraphics[width=\linewidth]{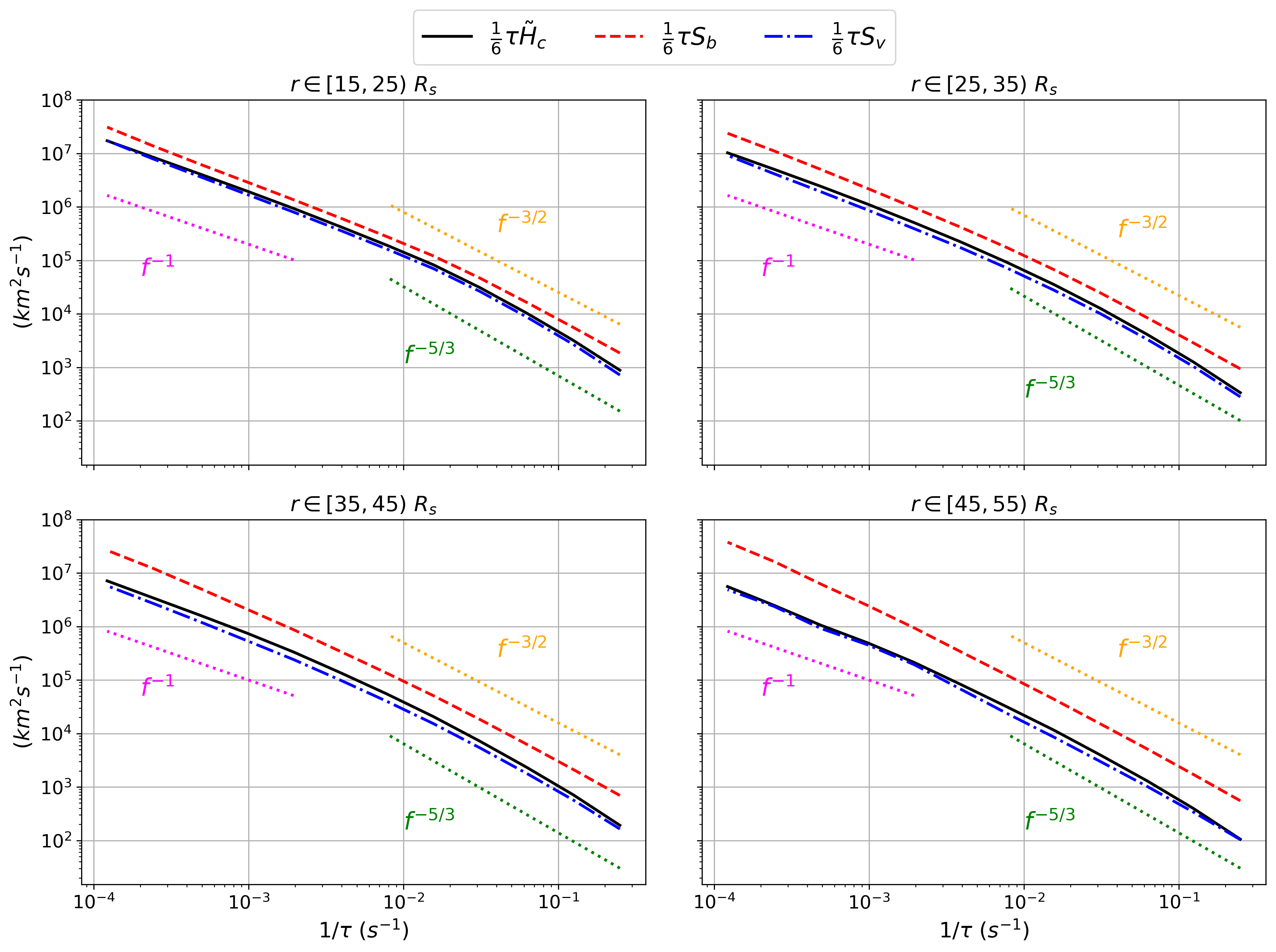}
  \caption{Equivalent spectra of rectified cross helicity (black solid curves) and second-order structure functions of magnetic field increments (red dashed curves), and velocity increments (blue dash-dotted curves) as functions of the reciprocal of the lag $\tau$. Each panel displays these for a different range of radial distance from the Sun. The dotted lines indicate spectral slopes with a power law index of -1 (pink), -3/2 (orange), and -5/3 (green).}
  \label{fig:spectra_at_different_r}
\end{figure*}

\subsection{Radial and Scale Dependence of  Alfvénicity \label{subsec: scale-dependence Alfvénicity}}

Figure \ref{fig:sigmac_vs_sigmar} shows scatter plots of the normalized cross helicity and the normalized residual energy during PSP solar encounters 8, 9 and 10 using the time lag $\tau = 87.4$ s, which is smaller than \add{the correlation time measured during E1 of 300 to 600 s \citep{Parashar2020},} in order to study Alfvénicity in the inertial range. 
Each 10-minute data point is colored according to the distance of PSP from the Sun.
According to Equation (\ref{eq:sigma-r-c-constraint}), $\sigma_c$ and $\sigma_r$ are constrained to lie within the unit circle.

At a small radial distance from the Sun, $|\sigma_c|$ tends to approach one, indicating the predominance of waves that propagate in one direction (either parallel or antiparallel to the mean field), with $\sigma_r$ tending towards zero according to the constraint of Equation (\ref{eq:sigma-r-c-constraint}). These properties are indicative of high Alfvénicity in the solar wind. 

However, as the heliocentric distance increases to $r\gtrsim0.35$ au, $|\sigma_c|$ decreases. $\sigma_r$ almost always exhibits a negative value. 
Indeed $\sigma_r$ is mostly distributed near the minimum value allowed by the constraint, and in some time intervals with $\sigma_c$ near 0, $\sigma_r$ is as low as -0.8 to -0.9 (though this is far from typical). Our results are consistent with previous results \citep{Chen2020,Shi_2021}, which indicate a decrease in $|\sigma_c|$ and $\sigma_r$ with increasing heliocentric distance $r$.

\begin{figure*}
  \includegraphics[width=\linewidth]{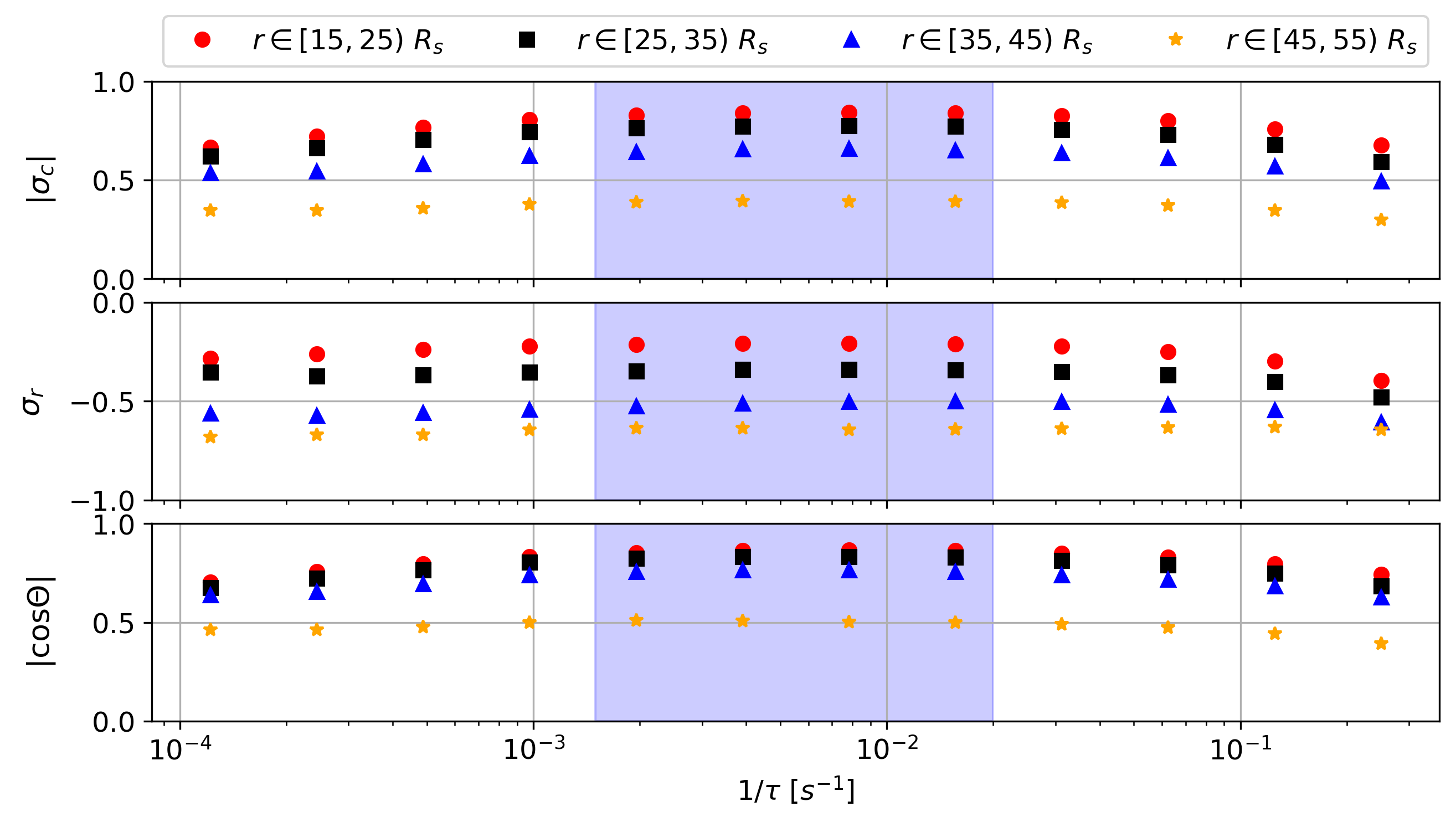}
  \caption{
  The magnitude of normalized cross helicity, normalized residual energy, and the magnitude of the global alignment cosine angle as a function of time lag over different ranges of distance from the Sun. 
  Each point indicates the mean value of the corresponding binned data. The blue-shaded region corresponds to the correlation scale uncertainty range \citep{Cuesta2022}.
  }
  \label{fig:scale_alf}
\end{figure*}

Figure \ref{fig:spectra_at_different_r} shows the equivalent spectra of rectified cross-helicity, $(1/6)\tau \tilde H_c$, and the second-order structure functions, $(1/6)\tau S_b$, and $(1/6)\tau S_v$, for PSP data from  different ranges of radial distance from the Sun. 
These spectra represent the average values of the second-order functions during encounters 8 to 10, with the Heliospheric Current Sheet (HCS) crossing events removed, plotted as functions of $1/\tau$, which can be interpreted as the fluctuation frequency $f$. 
Because previous work has reported an inertial range magnetic frequency spectrum proportional to $f^{-5/3}$ at larger $r$ and $f^{-3/2}$ at smaller $r$ in PSP data \citep{Chen2020}, we indicate these spectral dependences with dotted lines, as well as the $f^{-1}$ spectral dependence that has been found at lower frequencies \citep{Russell72,GoldsteinEA84}.

All of the equivalent spectra exhibit qualitatively similar trends, gradually but clearly bending from a steeper frequency dependence at high frequency to a flatter dependence at low frequency.
The frequency of transition from the inertial range to energy containing $f^{-1}$ range decreases with increasing heliocentric distance, which is consistent with the classic results from the Helios spacecraft \citep[e.g.,][]{BavassanoEA82,BrunoCarbone13,WuEA21}.
In the upper frequency range, corresponding to the inertial range of turbulence, our analysis is truncated near 1 NYs, the native cadence of the solar wind data from PSP/SWEAP.
Thus we analyze only about 1 order of magnitude of the inertial range and cannot precisely determine the power-law index of the frequency dependence, which seems consistent with a power-law index of either $-3/2$ or $-5/3$.
At lower frequency, all equivalent spectra for $r$ between 15 and $55 R_s$ are consistent with a gradual transition at decreasing frequency toward an $f^{-1}$ spectrum.  

Our results confirm that magnetic field energy dominates over kinetic energy in all heliospheric distance ranges and at all scales examined. Closer to the Sun, the difference between $S_b$ and $S_v$ magnitude is small; however, this difference increases greatly as the heliocentric distance increases.
Interestingly, $\tilde H_c$ is similar to $S_v$ at all distances at scales, but because $S_b$ becomes much greater than both of these with increasing $r$, the normalized cross helicity magnitude decreases greatly.

Figure \ref{fig:scale_alf} shows that $|\sigma_c|$, $\sigma_r$, and $|\cos\Theta|$, for which a higher value indicates stronger Alfv\'enicity, usually exhibit a dependence on the time lag, especially at distances close to the Sun. 
The blue-shaded region represents range of $\tau$ corresponding to the correlation length, where we employ the value obtained from \cite{Cuesta2022}.
Our analysis of Alfvénicity does not reach ion kinetic scales.

\begin{figure*}
  \centering
  \includegraphics[width=180mm]{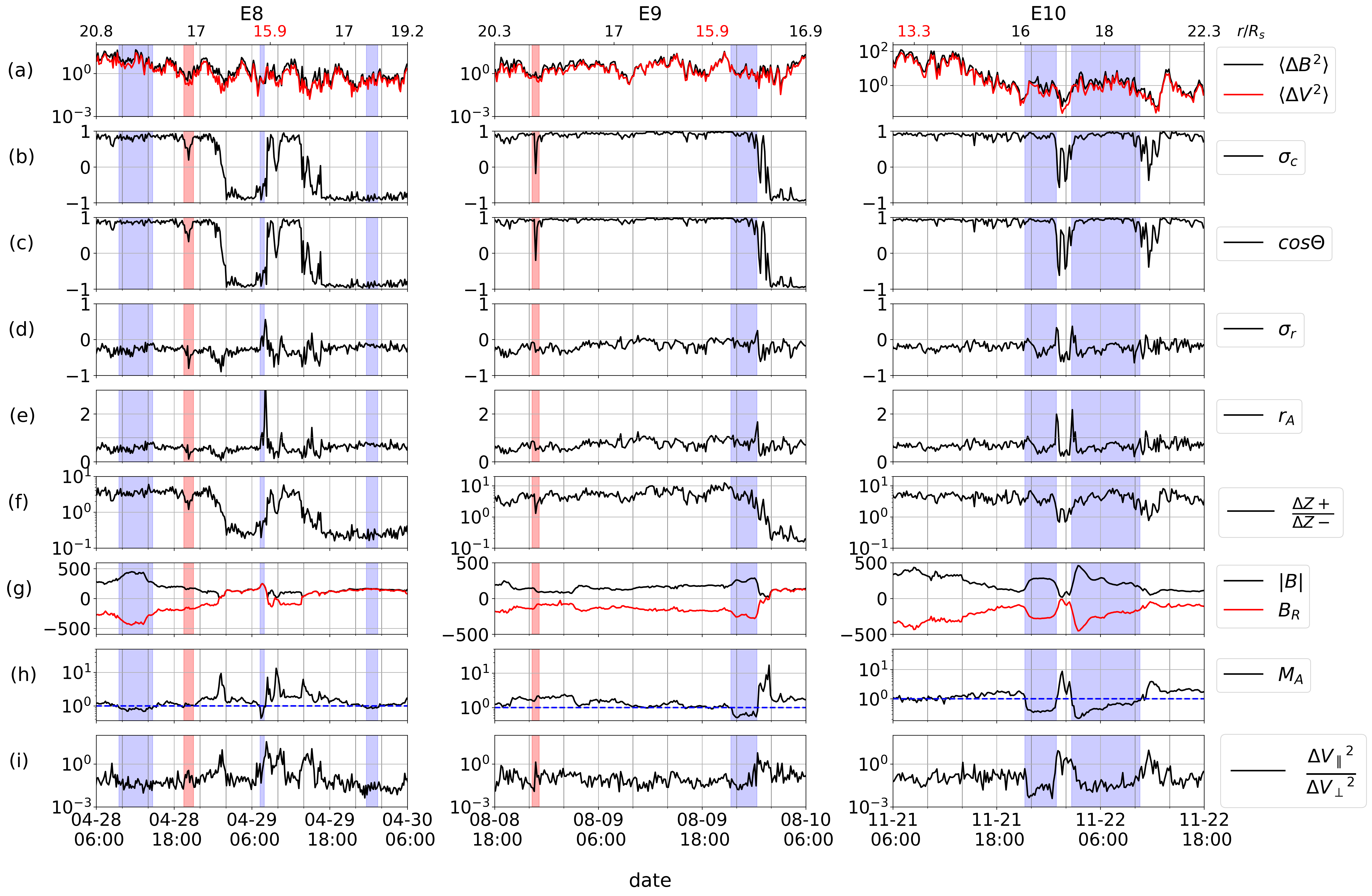}
  \caption{
  Time dependence of measures of Alfvénicity near the $8^{th}$ to $10^{th}$ PSP perihelia using \(\tau\) = 100 NYs (87.4 s). The top axes show the radial distance from the Sun in units of solar radii, with perihelion distance labeled in red.  The graphs show mean values of the (a) second-order structure functions of magnetic field and velocity, (b) normalized cross helicity, $\sigma_c$, (c) cosine of the global alignment angle between magnetic and velocity increments, $\cos{\Theta}$, (d) normalized residual energy, $\sigma_r$, (e) Alfvén ratio, $r_A$, (f) ratio of rms Elsässer amplitudes, (g) radial magnetic field component, $B_r$, and the magnetic field magnitude, $|B|$, (h) Alfvén Mach number, $M_A$, with a dashed blue line at $M_A = 1$, and (i) ratio of parallel to perpendicular velocity increments. Each panel shows a 10-minute average plotted at the centroid of the averaging interval. The blue shading indicates the sub-Alfvénic intervals with $M_A<1$ and red shading indicates the times of low Alfvénicity $|\sigma_c|<0.5$ other than full or partial crossings of the heliospheric current sheet.}
\label{fig:time_dependence}
\end{figure*}

We find that these measures of Alfvénicity
have their highest values for lags on the order of the correlation scale.
For $\tau$ below the correlation scale, $|\sigma_c|$, $\sigma_r$, and $|\cos\Theta|$ always decrease with decreasing $\tau$, indicating decreasing Alfv\'enicity.  
An exception is $\sigma_r$ in the highest distance range of 45 to $55R_s$, which fluctuates around a constant value and does not exhibit systematic scale dependence.
This trend of decreasing Alfvénicity as the lag decreases in the inertial range was previously reported by \cite{Parashar2018}, who observed this phenomenon in both the magnetosheath and the solar wind using data from the Magnetospheric Multiscale (MMS) spacecraft.

As the lag increases beyond the correlation length, the values of $|\sigma_c|$, $\sigma_r$, and $|\cos\Theta|$ again decrease, with the same exception as noted above. 
With further increases in the lag, there is evidently weaker correlation between increments of the velocity and magnetic fields. 

We also note that while all measures of Alfv\'enicity systematically decrease with increasing $r$, there is a sudden change in the scale-dependent cross helicity magnitude $|\sigma_c|$ and the alignment measure $|\cos\Theta|$ from the distance range of $[35,45)R_s$ to $[45,55)R_s$.
Furthermore, all measures of Alfv\'enicity shown in Figure 4 become nearly independent of scale at $r>45R_s$.


\subsection{Time Dependence of Alfvénicity}
 \label{subsec:time-dependence Alfvénicity}

Figure \ref{fig:time_dependence} illustrates the time dependence of Alfvénicity near perihelion during PSP solar encounters E8 to E10. 
In the context of the dependence on the radial distance described in Section 4.1, note that all these time periods have $r<25R_s$, corresponding to the innermost range of $r$ considered in that Section.
Here every parameter was calculated over non-overlapping 10-minute time windows.

Panel (a) displays the second-order structure functions from the velocity and magnetic increments. 
Throughout most of the time, we observed that $\langle\Delta B^2\rangle$ was slightly higher than $\langle\Delta V^2\rangle$, resulting in $\sigma_r<0$ and $r_A < 1$. 

Panels (b) and (c) show the normalized cross helicity and cosine of the global alignment angle, respectively, which exhibit very similar time series.
It is generally observed that Alfv\'enic fluctuations propagate predominantly outward from the Sun along magnetic field lines.  
Because a wave in ${\bf Z}^+={\bf V}+{\bf B}$ propagates antiparallel to the mean field,
then when $B_R<0$ (see panel (g)), the strongest increments are in $\Delta{\bf Z}^+$, with $\Delta {\bf V}\cdot\Delta{\bf B}>0$ and therefore $\sigma_c>0$ and $\cos\Theta>0$.
The predominance of $\Delta{\bf Z}^+$ can be seen from the generally high value of the ratio of rms values, $\Delta Z^+/\Delta Z^-\gg1$ when $B_R<0$ (see panel (f)).
On the other hand, when $B_R>0$, we generally observe stronger increments in $\Delta{\bf Z}^-$ with $\sigma_c<0$, $\cos\Theta<0$, and $\Delta Z^+/\Delta Z^-\ll1$.
A notable exception to this rule is that during a magnetic switchback, which can be defined as a temporary reversal in the $B_R$, the cross helicity is observed to be unchanged, as the predominant propagation direction of Alfv\'en waves continues to follow the field lines and reverses together with $B_R$ \citep{McManusEA20}.  
However, here we plot 10-minute averaged quantities and switchbacks are too small to be seen on this scale \citep{RaouafiEA23}.

While $\sigma_c$ and $\cos\Theta$ usually remain quite close to $\pm1$, the deviation from ideal Alfv\'enicity is more apparent when examining the normalized residual energy $\sigma_r$ and the Alfv\'en ratio $r_A$ (see panels (d) and (e), respectively).
For ideal Alfv\'enicity we would have $\sigma_r=0$ and $r_A=1$, but in these observations we almost always find $\sigma_r<0$ and $r_A<1$.


Sub-Alfvénic regions (highlighted in blue) are identified by $M_A<1$ (see panel (h)), where the Alfv\'enic Mach number is defined by $M_A\equiv V/V_A$. In the sub-Alfvénic solar wind, we typically observed high Alfvénicity and a low ratio of the mean squared parallel to perpendicular velocity increments (see panel (i)), which is consistent with the variance anisotropy in the sub-Alfvénic solar wind reported in \cite{BandyopadhyayEA22}. 
We observe no noticeable changes in Alfv\'enicity between sub-Alfv\'enic and super-Alfv\'enic solar wind over this range of heliospheric distance.



We observed upward spikes in $\sigma_r$ and $r_A$ at the edges of sub-Alfvénic regions at some but not all of the times when $\sigma_c$ and $\cos\Theta$ changed sign,
usually in association with PSP crossing boundaries between magnetic structures with different polarities or at full or partial HCS crossings.
Some of these events involved sharply higher electron density, resulting in lower $V_A$ and higher $M_A$, and if the wind outside the high-density structure was sub-Alfv\'enic, the sharply higher $M_A$ inside the structure caused the wind there to be super-Alfvenic.  
Hence some of these decreases in $\sigma_c$ and $\cos\Theta$, in some cases with upward spikes in $\sigma_r$ and $r_A$, occurred at the boundaries of sub-Alfv\'enic time periods.



\begin{figure*}
  \includegraphics[width=\linewidth]{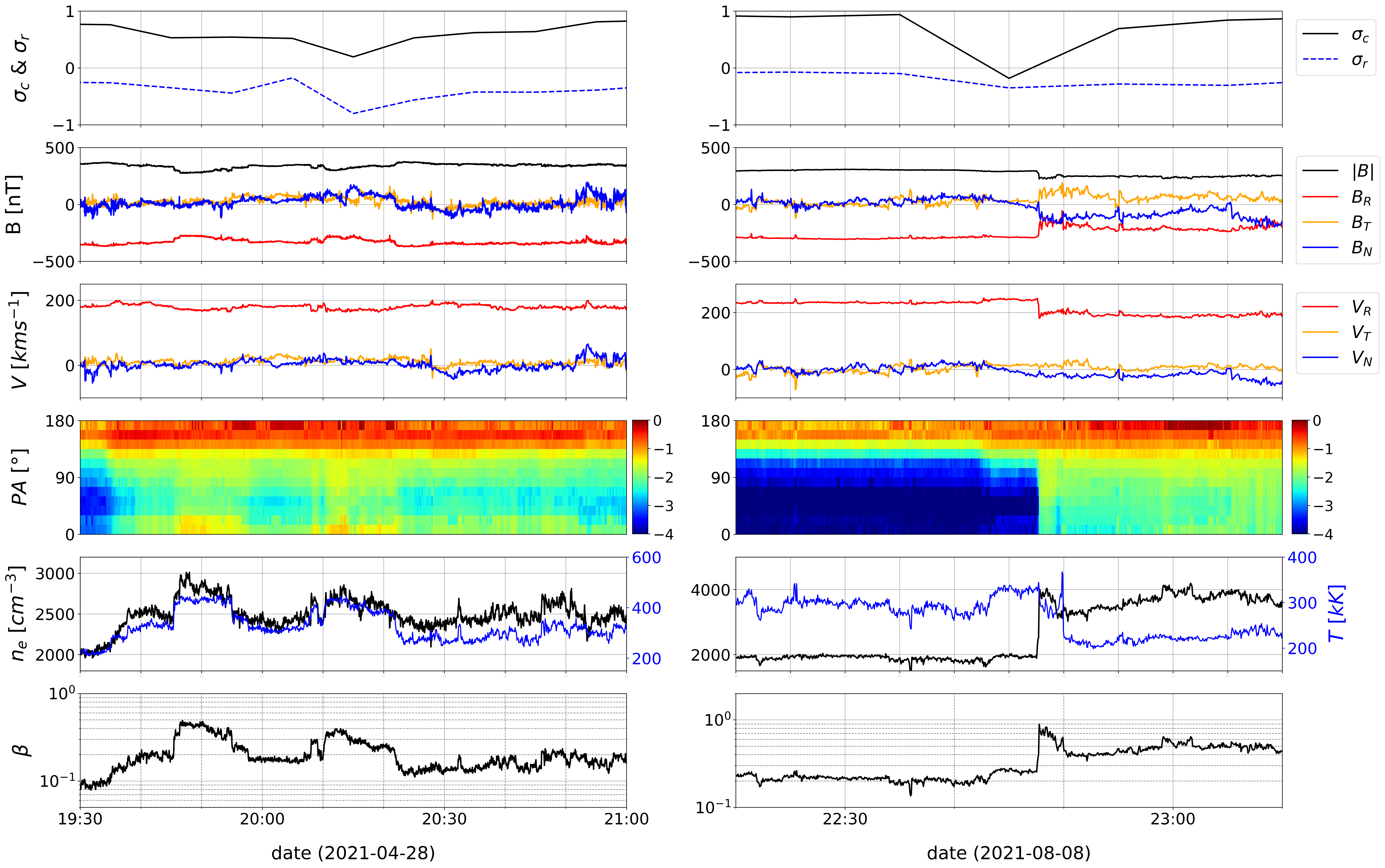}
  \caption{Zoom-in of the red-shaded time intervals in Figure \ref{fig:time_dependence}. 
From the top, the panels show the 10-minute averaged normalized cross helicity $\sigma_c$ and normalized residual energy $\sigma_r$ from increments over a lag of $\tau=$ 87.4 s, magnetic field components and magnitude, solar wind velocity components, pitch-angle distribution of 314 eV superthermal (strahl) electrons normalized by the maximum electron flux during the time of interest, electron density $n_e$ with ion temperature $T$, and plasma beta $\beta$.}
  \label{fig:low_alfvenic}
\end{figure*}

We consider the time intervals where $|\sigma_c| < 0.5$ for a time increment of $\tau = 87.4$ s to have low Alfvénicity. 
This definition allows us to identify and study the brief periods of time when the normalized cross helicity of the solar wind plasma is relatively low, indicating that the velocity and magnetic field fluctuations are not strongly correlated. 
By analyzing the characteristics of the solar wind during these periods, we can gain insights into the processes that are driving the anomalously low Alfvénicity, i.e., the factors that are contributing to the weaker correlation between the velocity and magnetic field fluctuations.

In contrast to the usual behavior that $\Delta Z^+/\Delta Z^-\gg1$ when $B_R<0$ and $\Delta Z^+/\Delta Z^-\ll1$ when $B_R>0$, 
during these periods of low Alfvénicity we observed a mixture of inward and outward-propagating waves, as indicated by $\Delta{Z^+}/\Delta{Z^-}\approx 1$.
Low Alfvénicity is commonly observed near the HCS. 
The times when PSP crosses the HCS are marked by a change in the polarity of the radial magnetic field $B_r$ from one polarity to another, corresponding with the change in degree of the suprathermal electron pitch-angle distribution from $0^{\circ}$ to $180^{\circ}$. Within the plasma sheet that surrounds the HCS, the properties of solar wind turbulence are  different from the normal solar wind. There is an increase in ion density, temperature, and flow speed, accompanied by a weakening of the magnetic field strength \citep{Smith2001}.

A reduction in Alfvénicity is also observed during incomplete or partial crossings. 
In these partial crossings, PSP approaches the HCS without crossing to the other side, as indicated by $B_r$ approaching zero but not fully reversing to the opposite direction, and 
there is a smaller magnetic field rotation compared with a complete crossing \citep{Phan_2021}.
The plot from Encounter 10 illustrates one such partial crossing occurring between two sub-Alfvénic regions as reported by \cite{Zhao_2022}. 
When PSP encounters such structures, we also observed that the velocity field is dominated by the parallel velocity component, unlike the highly Alfvénic solar wind.
  



Away from HCS crossings, we observed two additional time intervals near perihelia that exhibited low Alfvénicity, as indicated by the red shading in Figure \ref{fig:time_dependence}. 
In Figure \ref{fig:low_alfvenic} we show expanded plots for those brief non-Alfvénic time periods, showing $\sigma_c$, $\sigma_r$, and various solar wind parameters. 
Note that vector components are expressed in radial-tangential-normal (RTN) coordinates.
Both intervals exhibited $\sigma_c$ approaching zero, which indicates counter-propagating waves during this interval, and negative $\sigma_r$.

During the first interval of interest (19:30 to 21:00 UTC on 2021 April 28), the magnitude of the magnetic field was not constant and was accompanied by fluctuations in $n_e$, $T$, and $\beta$. The variability of these parameters suggests incompatibility with the Alfvén wave equation. Additionally, we observed a bi-directional electron pitch angle distribution. In the second interval of low Alfvénicity (22:20-23:10 UTC on 2021 August 8), we observed sudden changes in all magnetic field components, $V_R$, $n_e$, $T$, and $\beta$. 

\section{Discussion and Summary} \label{sec:discussion}


In this work, we study the scale-dependent Alfv\'enicity of the solar wind using recent data during PSP orbits 8 to 10 as it approached to within 14 $R_s$ of the Sun, in both super- and sub-Alfv\'enic solar wind.
The PSP data were analyzed using increment-based measures to investigate Alfv\'enic properties. This method yields Alfv\'enicity measures that exhibit smooth changes as a function of the increment time lag $\tau$. 


A structure function or related second-order function of increments at the time lag $\tau$ contains energy from all time scales smaller than and up to $\tau$. 
The derivative of the function with respect to $\tau$ expresses the energy at each time scale or frequency, so we can use an equivalent power spectrum based on that function that serves a similar purpose to the Fourier power spectrum. The equivalent spectrum exhibits a power-law break at the frequency marking the start (outer scale) of the inertial range of turbulence.
We confirm the classic result that the break frequency decreases with increasing distance from the Sun \citep[e.g.,][]{BavassanoEA82,BrunoCarbone13,WuEA21}.


The decreases in $|\sigma_c|$ and $|\cos\Theta|$ with increasing $r$ (see Figures \ref{fig:sigmac_vs_sigmar} and \ref{fig:scale_alf})
suggest that the magnitudes and directions of changes in the velocity $\Delta{\bf V}$ and the magnetic field $\Delta{\bf B}$ differ more strongly as the solar wind evolves. 
Indeed, all of our measures based on second-order functions of increments
indicate that Alfv\'enicity decreases with increasing heliocentric distance and at small scales, 
which is in agreement with previous studies using Fourier analysis \citep{Chen2020,Parashar2020,Shi_2021}. 




The Alfv\'en ratio $r_A$ and normalized residual energy $\sigma_r$ are additional indicators of Alfv\'enicity, where an ensemble of unidirectional Alfv\'en waves would have $r_A=1$ and $\sigma_r=0$. However, solar wind observations usually indicate that $r_A< 1$ and $\sigma_r<0$. Indeed, our Figure \ref{fig:sigmac_vs_sigmar} as well as many studies in the past \citep[e.g.,][]{BavassanoEA98,BrunoEA07} indicate that $\sigma_r$ in the solar wind is found near the minimum value compatible with the constraint $\sigma_c^2+\sigma_r^2\leq1$ (marked by the bounding circle in the Figure).
When the normalized cross-helicity is $\sigma_c=\pm1$, $\sigma_r$ is constrained to be zero, but otherwise it is usually found at lower (minimal) values. 
Negative values of $\sigma_r$ indicate that the cumulative kinetic energy of the solar wind at all scales less than the lag is lower than the magnetic fluctuation energy in the same range of scales. 
The predominance of magnetic fluctuation energy could be related to interference between inward and outward-type fluctuations, where the inward-type  can be generated by reflection in the expanding solar wind \citep{ZhouMatt89}. 
[However, it is interesting to note that a mixture of waves with mixed propagation directions (mixed cross helicities) is expected to have $r_A =1$ and $\sigma_r=0$ when the oppositely propagating wave packets are uncorrelated; this property is known as the ``Alfv\'en effect'' \citep{Kraichnan65, FyfeEA77b}.] 
In simulations, negative residual energy is generated through nonlinear interactions of counterpropagating fluctuations, i.e., ${\bf z^+}$ and ${\bf z^-}$ \citep{Shi_2023}. 
One physical mechanism for this is that current sheets generated in turbulent reconnection contribute to an excess of magnetic energy in the inertial range \citep{MatthaeusLamkin86}. 

The reader may note that in Figure \ref{fig:sigmac_vs_sigmar}, while the data points cluster near the circular boundary defined by the mathematical constraint $\sigma_c^2+\sigma_r^2\leq1$, there is a gap between the data points and that circle, with the exception of some data points at $\sigma_c=\pm1$ and $\sigma_r=0$.
In other words, while the data cluster near the minimum $\sigma_r$ for a given $\sigma_c$ as noted above, they do not reach the actual minimum value (except when $\sigma_c=\pm1$).
To understand the gap between the data and the boundary circle, let us consider the requirements to lie on the boundary circle. 
In general, $\sigma_c = \cos{\Theta} \sqrt{1-{\sigma_r}^2}$ (Equation [\ref{eq;sigma_c2}]), so reaching the boundary circle at $|\sigma_c|=\sqrt{1-{\sigma_r}^2}$ requires $|\cos\Theta|=1$, i.e., for each $t$ during the 10-minute averaging time there must be perfect directional alignment or anti-alignment, which implies that $\Delta {\bf V} = \alpha(t)\Delta {\bf B}$ for a scalar $\alpha(t)$.
When this occurs, this is a corollary of the local ``Beltrami'' alignment of magnetic and velocity fluctuations
\citep{TingEA86,StriblingMatthaeus91,MatthaeusEA08, ServidioEA08,OsmanEA11}.
Furthermore, $\alpha$ must take the same value at each $t$ within the interval. 
To see this, note that for any distribution of $\alpha$, the variance is given by
\begin{equation}
\langle(\alpha-\langle\alpha\rangle)^2\rangle 
= \langle\alpha^2\rangle-\langle\alpha\rangle^2 \geq 0
\end{equation}
which implies that 
\begin{equation}
\langle\alpha\rangle^2\leq\langle\alpha^2\rangle
\end{equation} and 
this is an equality if and only if the variance of $\alpha$ is zero.
Then we note that in this case of perfect directional alignment, we have 
\begin{eqnarray}
\sigma_c=\frac{2\langle\alpha\rangle}{\langle\alpha^2\rangle+1},&\phantom{=}&\quad\sigma_r=\frac{\langle\alpha^2\rangle-1}{\langle\alpha^2\rangle+1}\\
1-\sigma_c^2-\sigma_r^2&=&\frac{4\langle\alpha^2\rangle-4\langle\alpha\rangle^2}{(\langle\alpha^2\rangle+1)^2}
\end{eqnarray}
This is non-negative, and the data point lies on the boundary circle if and only if $\langle\alpha\rangle^2=\langle\alpha^2\rangle$ and the variance of $\alpha(t)$ is zero within that interval, so $\alpha$ is constant and $r_A=1/\alpha^2$.
In other words, if we do not have ideal Alfv\'enicity ($|\sigma_c|\ne1$), then $\Delta V$ deviates from $\pm\Delta B$, but the only way for a data point to remain on the boundary circle is to deviate to the same fixed fraction $\alpha$, i.e., $\Delta {\bf V} = \alpha\Delta {\bf B}$, at every $t$ value.
 
Thus we have a physical picture that near the Sun there is often near-ideal Alfv\'enicity with $\sigma_c\approx1$ and $\sigma_r\approx0$, as we see from Figure \ref{fig:sigmac_vs_sigmar}, which is consistent with the fluctuations being mostly Alfv\'en waves generated by the Sun. 
As the solar wind evolves farther from the Sun, the processes (interactions) that cause a deviation from ideal Alfv\'enicity, so that $\sigma_c\ne\pm1$, also cause sufficient variation that we do not obtain increments of the {\it same} direction and magnitude ($\Delta {\bf V} = \alpha\Delta {\bf B}$) at every $t$ value.  This leads to the zone of avoidance at the boundary where $\sigma_c^2+\sigma_r^2=1$, despite the tendency noted above for $\sigma_r$ to approach near-minimal values.

Our results for $|\sigma_c|$ and $\sigma_r$ during E8 to E10 exhibit lower values compared with those reported in a study of E1 by \cite{Chen2020}, even when comparing at the same radial distance. 
According to \citet{Shi_2021}, Alfv\'enicity is weaker in slower solar wind, so the difference between our results and those of \citet{Chen2020} could be attributed to the lower solar wind speeds observed during E8 to E10 perihelia in comparison with E1; \citet{Shi_2021} reported a similar difference when examining data from E5, which also exhibited a slower solar wind speed than E1.

Near the Alfvén critical zone, with the radial velocity exceeding the Alfvén speed, the flow cannot propagate backward to the Sun \citep{DeForest2014}. It is interesting to look for possible changes in Alfvénicity when PSP crosses this zone. 
However, we did not observe noticeable changes in Alfvénicity parameters between sub-Alfv\'enic and super-Alfv\'enic wind in PSP data, which indeed is consistent with the model results of \citet{LouEA1994}. 

We find that scale-dependent Afv\'enicity peaks at $\tau$ comparable to the correlation time and usually decreases at other $\tau$ ranges significantly different from the correlation time. 
In turbulence the smaller scales of the inertial range have a shorter time scale for nonlinear interactions that tend to disrupt the Alfv\'enic correlations and reduce Alfv\'enicity.
Shear was also proposed by \cite{Parashar2020} to serve as a factor responsible for reducing $|\cos\Theta|$
\citep[see also][]{RobertsEA87,RobertsEA1992,ZankEA96}. 
Without shear, $|\cos\Theta|$ would be expected to increase at $\tau$ smaller than the correlation time \citep{boldyrev2006}, as verified by computer simulations \citep{mason2006}. 

As $\tau$ increases above the correlation scale, the magnetic field structure function rises due to the increment technique acting as a high-pass filter in the frequency $1/\tau$, 
accumulating the contribution of energy from fluctuations at scales smaller than $\tau$. As a result, the structure functions increase with increasing $\tau$. However, the relationship between magnetic and velocity fluctuations is more limited to the scales comparable to the correlation time. Beyond the correlation time ($\tau$), we find that Alfv\'enicity becomes weaker. 
For example, as noted earlier the normalized cross helicity is the ratio between the cross helicity $H_c$ and the energy $E=(1/2)(S_v+S_b)$.
Thus the gradual decrease in $|\sigma_c|$ with increasing lag $\tau$ can be understood as $H_c$ increasing more slowly than $E$, as can be seen visually from Figures \ref{fig:Structure} and \ref{fig:spectra_at_different_r}.
We note that some previous studies in the past defined increments in terms of fluctuating fields, relative to a running average, whereas here we use the ${\bf V}$ and ${\bf B}$ fields directly.  
Subtracting a running average does tend to remove the effects of fluctuations longer than the averaging time, and may lead to different behaviors at long $\tau$.
The decrease in $|\sigma_c|$ that we observe as $\tau$ increases beyond the correlation length
is pronounced at closer distances to the Sun, but as we move further from the Sun, the Alfvénicity trend at different $\tau$ values becomes flatter.  

The heliospheric plasma sheet (HPS) is a broader region characterized by high density and high $\beta$ surrounding the HCS, which may have repeated structures of flux ropes and magnetic reconnection sites \citep{Sanchez-DiazEA19}.
The values of $\sigma_c$ and $\cos\Theta$ approach or cross zero as PSP nears the HCS, presumably because the solar wind is affected by the HPS.
This behavior indicates the presence of non-Alfvénic fluctuations and a reduced degree of alignment (or anti-alignment) between velocity and magnetic field increments. 
The decrease of $|\sigma_c|$ is also associated with strong velocity shear that occurs close to the current sheet \citep{RobertsEA1992}. 
Around the HPS, the inward and outward propagating waves are strongly mixed. These observations can be attributed to the differences in the mean magnetic field across the HCS and the absence of correlation between fields from opposite sides of the sheet. 

During times when Alfv\'enicity is observed to be particularly low, PSP was crossing a boundary between regions with different plasma properties, as found in the HPS, velocity shear structures, and magnetic flux tube boundaries, with scattered strahl electrons  in a non-unidirectional distribution (see Figure \ref{fig:low_alfvenic}). 
These structures can scatter Alfv\'en waves, causing interactions between oppositely propagating wave packets and disrupting the correlation between velocity and magnetic field fluctuations, resulting in lower Alfvénicity. 
More specifically, if sum of the parallel wavenumbers of the interacting waves is near zero, the interactions tend to produce two-dimensional fluctuations with wavenumbers perpendicular to the mean field \citep{ShebalinEA83,Tu_Marsch_1993}, with predominantly magnetic rather than kinetic energy, tending to make the residual energy more negative as well \citep{WangEA11,HowesNielson13}. 
%
The decrease in Alfvénicity during these times may also be attributed to the spacecraft crossing discontinuity such as the boundary of a flux tube, where a random value of $\sigma_c$ is encountered. 

This research is funded by Kasetsart University through the Graduate School Fellowship Program. It has been partially supported in Thailand by Thailand Science Research and Innovation (RTA6280002), by the National Science and Technology Development Agency (NSTDA) and National Research Council of Thailand (NRCT): High-Potential Research Team Grant Program (N42A650868), 
and from the NSRF via the Program Management Unit for Human Resources \& Institutional Development, Research and Innovation (B37G660015). It was also supported by the Parker Solar Probe mission under the ISOIS project (contract NNN06AA01C) and a subcontract to University of Delaware from Princeton University. R.K. acknowledges the support of the Centre National de la Recherche Scientique (CNRS), the University of Toulouse III (UPS), and the Centre National d’Études Spatiales (CNES), France.


\bibliography{alfvenicity_refs}{}

\begin{thebibliography}{}
\expandafter\ifx\csname natexlab\endcsname\relax\def\natexlab#1{#1}\fi
\providecommand{\url}[1]{\href{#1}{#1}}
\providecommand{\dodoi}[1]{doi:~\href{http://doi.org/#1}{\nolinkurl{#1}}}
\providecommand{\doeprint}[1]{\href{http://ascl.net/#1}{\nolinkurl{http://ascl.net/#1}}}
\providecommand{\doarXiv}[1]{\href{https://arxiv.org/abs/#1}{\nolinkurl{https://arxiv.org/abs/#1}}}

\bibitem[{Bale {et~al.}(2016)Bale, Goetz, Harvey, Turin, Bonnell, de~Wit,
  Ergun, MacDowall, Pulupa, Andre, Bolton, Bougeret, Bowen, Burgess, Cattell,
  Chandran, Chaston, Chen, Choi, Connerney, Cranmer, Diaz-Aguado, Donakowski,
  Drake, Farrell, Fergeau, Fermin, Fischer, Fox, Glaser, Goldstein, Gordon,
  Hanson, Harris, Hayes, Hinze, Hollweg, Horbury, Howard, Hoxie, Jannet,
  Karlsson, Kasper, Kellogg, Kien, Klimchuk, Krasnoselskikh, Krucker, Lynch,
  Maksimovic, Malaspina, Marker, Martin, Martinez-Oliveros, McCauley, McComas,
  McDonald, Meyer-Vernet, Moncuquet, Monson, Mozer, Murphy, Odom, Oliverson,
  Olson, Parker, Pankow, Phan, Quataert, Quinn, Ruplin, Salem, Seitz, Sheppard,
  Siy, Stevens, Summers, Szabo, Timofeeva, Vaivads, Velli, Yehle, Werthimer, \&
  Wygant}]{Bale_2016}
Bale, S.~D., Goetz, K., Harvey, P.~R., {et~al.} 2016, Space science reviews,
  204, 49–82, \dodoi{10.1007/s11214-016-0244-5}

\bibitem[{{Bandyopadhyay} {et~al.}(2022){Bandyopadhyay}, {Matthaeus},
  {McComas}, {Chhiber}, {Usmanov}, {Huang}, {Livi}, {Larson}, {Kasper}, {Case},
  {Stevens}, {Whittlesey}, {Romeo}, {Bale}, {Bonnell}, {Dudok de Wit}, {Goetz},
  {Harvey}, {MacDowall}, {Malaspina}, \& {Pulupa}}]{BandyopadhyayEA22}
{Bandyopadhyay}, R., {Matthaeus}, W.~H., {McComas}, D.~J., {et~al.} 2022,
  \apjl, 926, L1, \dodoi{10.3847/2041-8213/ac4a5c}

\bibitem[{{Barnes}(1979)}]{Barnes79}
{Barnes}, A. 1979, in Solar System Plasma Physics, ed. E.~N. {Parker}, C.~F.
  {Kennel}, \& L.~J. {Lanzerotti}, Vol.~1, 249--319

\bibitem[{{Barnes}(1981)}]{Barnes81}
---. 1981, \jgr, 86, 7498, \dodoi{10.1029/JA086iA09p07498}

\bibitem[{{Bavassano} {et~al.}(1982){Bavassano}, {Dobrowolny}, {Mariani}, \&
  {Ness}}]{BavassanoEA82}
{Bavassano}, B., {Dobrowolny}, M., {Mariani}, F., \& {Ness}, N.~F. 1982, \jgr,
  87, 3617, \dodoi{10.1029/JA087iA05p03617}

\bibitem[{{Bavassano} {et~al.}(1998){Bavassano}, {Pietropaolo}, \&
  {Bruno}}]{BavassanoEA98}
{Bavassano}, B., {Pietropaolo}, E., \& {Bruno}, R. 1998, \jgr, 103, 6521,
  \dodoi{10.1029/97JA03029}

\bibitem[{Belcher \& Davis(1971)}]{Belcher_Davis_1971}
Belcher, J.~W., \& Davis, Leverett, J. 1971, Journal of geophysical research,
  76, 3534–3563, \dodoi{10.1029/ja076i016p03534}

\bibitem[{{Boldyrev}(2006)}]{boldyrev2006}
{Boldyrev}, S. 2006, \prl, 96, 115002, \dodoi{10.1103/PhysRevLett.96.115002}

\bibitem[{{Bruno} {et~al.}(1985){Bruno}, {Bavassano}, \&
  {Villante}}]{BrunoEA85}
{Bruno}, R., {Bavassano}, B., \& {Villante}, U. 1985, \jgr, 90, 4373,
  \dodoi{10.1029/JA090iA05p04373}

\bibitem[{{Bruno} \& {Carbone}(2013)}]{BrunoCarbone13}
{Bruno}, R., \& {Carbone}, V. 2013, Living Reviews in Solar Physics, 10, 2,
  \dodoi{10.12942/lrsp-2013-2}

\bibitem[{Bruno {et~al.}(2003)Bruno, Carbone, Sorriso-Valvo, \&
  Bavassano}]{Bruno_2003}
Bruno, R., Carbone, V., Sorriso-Valvo, L., \& Bavassano, B. 2003, Journal of
  Geophysical Research: Space Physics, 108,
  \dodoi{https://doi.org/10.1029/2002JA009615}

\bibitem[{{Bruno} {et~al.}(2007){Bruno}, {D'Amicis}, {Bavassano}, {Carbone}, \&
  {Sorriso-Valvo}}]{BrunoEA07}
{Bruno}, R., {D'Amicis}, R., {Bavassano}, B., {Carbone}, V., \&
  {Sorriso-Valvo}, L. 2007, Annales Geophysicae, 25, 1913,
  \dodoi{10.5194/angeo-25-1913-2007}

\bibitem[{{Chasapis} {et~al.}(2017){Chasapis}, {Matthaeus}, {Parashar},
  {Fuselier}, {Maruca}, {Phan}, {Burch}, {Moore}, {Pollock}, {Gershman},
  {Torbert}, {Russell}, \& {Strangeway}}]{ChasapisEA17}
{Chasapis}, A., {Matthaeus}, W.~H., {Parashar}, T.~N., {et~al.} 2017, \apjl,
  844, L9, \dodoi{10.3847/2041-8213/aa7ddd}

\bibitem[{{Chen} {et~al.}(2020){Chen}, {Bale}, {Bonnell}, {Borovikov}, {Bowen},
  {Burgess}, {Case}, {Chandran}, {de Wit}, {Goetz}, {Harvey}, {Kasper},
  {Klein}, {Korreck}, {Larson}, {Livi}, {MacDowall}, {Malaspina}, {Mallet},
  {McManus}, {Moncuquet}, {Pulupa}, {Stevens}, \& {Whittlesey}}]{Chen2020}
{Chen}, C.~H.~K., {Bale}, S.~D., {Bonnell}, J.~W., {et~al.} 2020, \apjs, 246,
  53, \dodoi{10.3847/1538-4365/ab60a3}

\bibitem[{{Chhiber} {et~al.}(2022){Chhiber}, {Matthaeus}, {Usmanov},
  {Bandyopadhyay}, \& {Goldstein}}]{ChhiberEA22}
{Chhiber}, R., {Matthaeus}, W.~H., {Usmanov}, A.~V., {Bandyopadhyay}, R., \&
  {Goldstein}, M.~L. 2022, \mnras, 513, 159, \dodoi{10.1093/mnras/stac779}

\bibitem[{{Chhiber} {et~al.}(2018){Chhiber}, {Chasapis}, {Bandyopadhyay},
  {Parashar}, {Matthaeus}, {Maruca}, {Moore}, {Burch}, {Torbert}, {Russell},
  {Le Contel}, {Argall}, {Fischer}, {Mirioni}, {Strangeway}, {Pollock},
  {Giles}, \& {Gershman}}]{ChhiberEA2018}
{Chhiber}, R., {Chasapis}, A., {Bandyopadhyay}, R., {et~al.} 2018, Journal of
  Geophysical Research (Space Physics), 123, 9941, \dodoi{10.1029/2018JA025768}

\bibitem[{{Cuesta} {et~al.}(2022){Cuesta}, {Chhiber}, {Roy}, {Goodwill},
  {Pecora}, {Jarosik}, {Matthaeus}, {Parashar}, \&
  {Bandyopadhyay}}]{Cuesta2022}
{Cuesta}, M.~E., {Chhiber}, R., {Roy}, S., {et~al.} 2022, \apjl, 932, L11,
  \dodoi{10.3847/2041-8213/ac73fd}

\bibitem[{{D'Amicis} {et~al.}(2021){D'Amicis}, {Perrone}, {Bruno}, \&
  {Velli}}]{DAmicisEA21}
{D'Amicis}, R., {Perrone}, D., {Bruno}, R., \& {Velli}, M. 2021, Journal of
  Geophysical Research (Space Physics), 126, e28996,
  \dodoi{10.1029/2020JA028996}

\bibitem[{{Davidson}(2004)}]{Davidson15}
{Davidson}, P.~A. 2004, {Turbulence : an introduction for scientists and
  engineers}

\bibitem[{{DeForest} {et~al.}(2018){DeForest}, {Howard}, {Velli}, {Viall}, \&
  {Vourlidas}}]{DeForestEA18}
{DeForest}, C.~E., {Howard}, R.~A., {Velli}, M., {Viall}, N., \& {Vourlidas},
  A. 2018, \apj, 862, 18, \dodoi{10.3847/1538-4357/aac8e3}

\bibitem[{{DeForest} {et~al.}(2014){DeForest}, {Howard}, \&
  {McComas}}]{DeForest2014}
{DeForest}, C.~E., {Howard}, T.~A., \& {McComas}, D.~J. 2014, \apj, 787, 124,
  \dodoi{10.1088/0004-637X/787/2/124}

\bibitem[{Fox {et~al.}(2016)Fox, Velli, Bale, Decker, Driesman, Howard, Kasper,
  Kinnison, Kusterer, Lario, Lockwood, McComas, Raouafi, \& Szabo}]{Fox_2016}
Fox, N.~J., Velli, M.~C., Bale, S.~D., {et~al.} 2016, Space science reviews,
  204, 7–48, \dodoi{10.1007/s11214-015-0211-6}

\bibitem[{{Frisch}(1995)}]{Frisch95}
{Frisch}, U. 1995, {Turbulence. The legacy of A.N. Kolmogorov}

\bibitem[{Fyfe {et~al.}(1977)Fyfe, Montgomery, \& Joyce}]{FyfeEA77b}
Fyfe, D., Montgomery, D., \& Joyce, G. 1977, 17, 369

\bibitem[{{Goldstein} {et~al.}(1984){Goldstein}, {Burlaga}, \&
  {Matthaeus}}]{GoldsteinEA84}
{Goldstein}, M.~L., {Burlaga}, L.~F., \& {Matthaeus}, W.~H. 1984, \jgr, 89,
  3747, \dodoi{10.1029/JA089iA06p03747}

\bibitem[{{Goldstein} {et~al.}(1974){Goldstein}, {Klimas}, \&
  {Barish}}]{Goldstein1974}
{Goldstein}, M.~L., {Klimas}, A.~J., \& {Barish}, F.~D. 1974, in Solar Wind
  Three, ed. C.~T. {Russell}, 385--387

\bibitem[{{Howes} \& {Nielson}(2013)}]{HowesNielson13}
{Howes}, G.~G., \& {Nielson}, K.~D. 2013, Physics of Plasmas, 20, 072302,
  \dodoi{10.1063/1.4812805}

\bibitem[{Kasper {et~al.}(2016)Kasper, Abiad, Austin, Balat-Pichelin, Bale,
  Belcher, Berg, Bergner, Berthomier, Bookbinder, Brodu, Caldwell, Case,
  Chandran, Cheimets, Cirtain, Cranmer, Curtis, Daigneau, Dalton, Dasgupta,
  DeTomaso, Diaz-Aguado, Djordjevic, Donaskowski, Effinger, Florinski, Fox,
  Freeman, Gallagher, Gary, Gauron, Gates, Goldstein, Golub, Gordon, Gurnee,
  Guth, Halekas, Hatch, Heerikuisen, Ho, Hu, Johnson, Jordan, Korreck, Larson,
  Lazarus, Li, Livi, Ludlam, Maksimovic, McFadden, Marchant, Maruca, McComas,
  Messina, Mercer, Park, Peddie, Pogorelov, Reinhart, Richardson, Robinson,
  Rosen, Skoug, Slagle, Steinberg, Stevens, Szabo, Taylor, Tiu, Turin, Velli,
  Webb, Whittlesey, Wright, Wu, \& Zank}]{Kasper_2016}
Kasper, J.~C., Abiad, R., Austin, G., {et~al.} 2016, Space science reviews,
  204, 131–186, \dodoi{10.1007/s11214-015-0206-3}

\bibitem[{{Kasper} {et~al.}(2019){Kasper}, {Bale}, {Belcher}, {Berthomier},
  {Case}, {Chandran}, {Curtis}, {Gallagher}, {Gary}, {Golub}, {Halekas}, {Ho},
  {Horbury}, {Hu}, {Huang}, {Klein}, {Korreck}, {Larson}, {Livi}, {Maruca},
  {Lavraud}, {Louarn}, {Maksimovic}, {Martinovic}, {McGinnis}, {Pogorelov},
  {Richardson}, {Skoug}, {Steinberg}, {Stevens}, {Szabo}, {Velli},
  {Whittlesey}, {Wright}, {Zank}, {MacDowall}, {McComas}, {McNutt}, {Pulupa},
  {Raouafi}, \& {Schwadron}}]{KasperEA19}
{Kasper}, J.~C., {Bale}, S.~D., {Belcher}, J.~W., {et~al.} 2019, \nat, 576,
  228, \dodoi{10.1038/s41586-019-1813-z}

\bibitem[{{Kasper} {et~al.}(2021){Kasper}, {Klein}, {Lichko}, {Huang}, {Chen},
  {Badman}, {Bonnell}, {Whittlesey}, {Livi}, {Larson}, {Pulupa}, {Rahmati},
  {Stansby}, {Korreck}, {Stevens}, {Case}, {Bale}, {Maksimovic}, {Moncuquet},
  {Goetz}, {Halekas}, {Malaspina}, {Raouafi}, {Szabo}, {MacDowall}, {Velli},
  {Dudok de Wit}, \& {Zank}}]{KasperEA21}
{Kasper}, J.~C., {Klein}, K.~G., {Lichko}, E., {et~al.} 2021, \prl, 127,
  255101, \dodoi{10.1103/PhysRevLett.127.255101}

\bibitem[{{Kraichnan}(1965)}]{Kraichnan65}
{Kraichnan}, R.~H. 1965, Physics of Fluids, 8, 1385, \dodoi{10.1063/1.1761412}

\bibitem[{{Lindborg}(1999)}]{Lindborg99}
{Lindborg}, E. 1999, Journal of Fluid Mechanics, 388, 259,
  \dodoi{10.1017/S0022112099004851}

\bibitem[{{Lou}(1994)}]{LouEA1994}
{Lou}, Y.-Q. 1994, \jgr, 99, 14747, \dodoi{10.1029/94JA00928}

\bibitem[{{Mason} {et~al.}(2006){Mason}, {Cattaneo}, \& {Boldyrev}}]{mason2006}
{Mason}, J., {Cattaneo}, F., \& {Boldyrev}, S. 2006, \prl, 97, 255002,
  \dodoi{10.1103/PhysRevLett.97.255002}

\bibitem[{{Matthaeus} \& {Goldstein}(1982)}]{MatthaeusGoldstein82meas}
{Matthaeus}, W.~H., \& {Goldstein}, M.~L. 1982, \jgr, 87, 6011,
  \dodoi{10.1029/JA087iA08p06011}

\bibitem[{Matthaeus \& Lamkin(1986)}]{MatthaeusLamkin86}
Matthaeus, W.~H., \& Lamkin, S.~L. 1986, The Physics of fluids, 29, 2513,
  \dodoi{10.1063/1.866004}

\bibitem[{{Matthaeus} {et~al.}(2008){Matthaeus}, {Pouquet}, {Mininni},
  {Dmitruk}, \& {Breech}}]{MatthaeusEA08}
{Matthaeus}, W.~H., {Pouquet}, A., {Mininni}, P.~D., {Dmitruk}, P., \&
  {Breech}, B. 2008, \prl, 100, 085003, \dodoi{10.1103/PhysRevLett.100.085003}

\bibitem[{McComas {et~al.}(2000)McComas, Barraclough, Funsten, Gosling,
  Santiago-Muñoz, Skoug, Goldstein, Neugebauer, Riley, \&
  Balogh}]{McComas_2000}
McComas, D.~J., Barraclough, B.~L., Funsten, H.~O., {et~al.} 2000, Journal of
  geophysical research, 105, 10419–10433, \dodoi{10.1029/1999ja000383}

\bibitem[{{McManus} {et~al.}(2020){McManus}, {Bowen}, {Mallet}, {Chen},
  {Chandran}, {Bale}, {Larson}, {Dudok de Wit}, {Kasper}, {Stevens},
  {Whittlesey}, {Livi}, {Korreck}, {Goetz}, {Harvey}, {Pulupa}, {MacDowall},
  {Malaspina}, {Case}, \& {Bonnell}}]{McManusEA20}
{McManus}, M.~D., {Bowen}, T.~A., {Mallet}, A., {et~al.} 2020, \apjs, 246, 67,
  \dodoi{10.3847/1538-4365/ab6dce}

\bibitem[{Moncuquet {et~al.}(2020)Moncuquet, Meyer-Vernet, Issautier, Pulupa,
  Bonnell, Bale, de~Wit, Goetz, Griton, Harvey, MacDowall, Maksimovic, \&
  Malaspina}]{Moncuquet_2020}
Moncuquet, M., Meyer-Vernet, N., Issautier, K., {et~al.} 2020, The
  Astrophysical journal. Supplement series, 246, 44,
  \dodoi{10.3847/1538-4365/ab5a84}

\bibitem[{Monin \& Yaglom(1999)}]{MoninYaglom}
Monin, A.~S., \& Yaglom, A. 1999, Statistical fluid mechanics: The mechanics of
  turbulence, Tech. rep., Massachusetts Inst of Tech Cambridge

\bibitem[{{Osman} {et~al.}(2011){Osman}, {Wan}, {Matthaeus}, {Breech}, \&
  {Oughton}}]{OsmanEA11}
{Osman}, K.~T., {Wan}, M., {Matthaeus}, W.~H., {Breech}, B., \& {Oughton}, S.
  2011, \apj, 741, 75, \dodoi{10.1088/0004-637X/741/2/75}

\bibitem[{{Parashar} {et~al.}(2018){Parashar}, {Chasapis}, {Bandyopadhyay},
  {Chhiber}, {Matthaeus}, {Maruca}, {Shay}, {Burch}, {Moore}, {Giles},
  {Gershman}, {Pollock}, {Torbert}, {Russell}, {Strangeway}, \&
  {Roytershteyn}}]{Parashar2018}
{Parashar}, T.~N., {Chasapis}, A., {Bandyopadhyay}, R., {et~al.} 2018, \prl,
  121, 265101, \dodoi{10.1103/PhysRevLett.121.265101}

\bibitem[{Parashar {et~al.}(2020)Parashar, Goldstein, Maruca, Matthaeus,
  Ruffolo, Bandyopadhyay, Chhiber, Chasapis, Qudsi, Vech, Roberts, Bale,
  Bonnell, de~Wit, Goetz, Harvey, MacDowall, Malaspina, Pulupa, Kasper,
  Korreck, Case, Stevens, Whittlesey, Larson, Livi, Velli, \&
  Raouafi}]{Parashar2020}
Parashar, T.~N., Goldstein, M.~L., Maruca, B.~A., {et~al.} 2020, The
  Astrophysical journal. Supplement series, 246, 58,
  \dodoi{10.3847/1538-4365/ab64e6}

\bibitem[{Phan {et~al.}(2021)Phan, Lavraud, Halekas, Øieroset, Drake,
  Eastwood, Shay, Pyakurel, Bale, Larson, Livi, Whittlesey, Rahmati, Pulupa,
  McManus, Verniero, Bonnell, Schwadron, Stevens, Case, Kasper, MacDowall,
  Szabo, Koval, Korreck, Dudok~de Wit, Malaspina, Goetz, \& Harvey}]{Phan_2021}
Phan, T.~D., Lavraud, B., Halekas, J.~S., {et~al.} 2021, Astronomy and
  astrophysics, 650, A13, \dodoi{10.1051/0004-6361/202039863}

\bibitem[{{Raouafi} {et~al.}(2023){Raouafi}, {Matteini}, {Squire}, {Badman},
  {Velli}, {Klein}, {Chen}, {Matthaeus}, {Szabo}, {Linton}, {Allen}, {Szalay},
  {Bruno}, {Decker}, {Akhavan-Tafti}, {Agapitov}, {Bale}, {Bandyopadhyay},
  {Battams}, {Ber{\v{c}}i{\v{c}}}, {Bourouaine}, {Bowen}, {Cattell},
  {Chandran}, {Chhiber}, {Cohen}, {D'Amicis}, {Giacalone}, {Hess}, {Howard},
  {Horbury}, {Jagarlamudi}, {Joyce}, {Kasper}, {Kinnison}, {Laker}, {Liewer},
  {Malaspina}, {Mann}, {McComas}, {Niembro-Hernandez}, {Nieves-Chinchilla},
  {Panasenco}, {Pokorn{\'y}}, {Pusack}, {Pulupa}, {Perez}, {Riley},
  {Rouillard}, {Shi}, {Stenborg}, {Tenerani}, {Verniero}, {Viall}, {Vourlidas},
  {Wood}, {Woodham}, \& {Woolley}}]{RaouafiEA23}
{Raouafi}, N.~E., {Matteini}, L., {Squire}, J., {et~al.} 2023, \ssr, 219, 8,
  \dodoi{10.1007/s11214-023-00952-4}

\bibitem[{{Roberts} {et~al.}(1987){Roberts}, {Goldstein}, {Klein}, \&
  {Matthaeus}}]{RobertsEA87}
{Roberts}, D.~A., {Goldstein}, M.~L., {Klein}, L.~W., \& {Matthaeus}, W.~H.
  1987, \jgr, 92, 12023, \dodoi{10.1029/JA092iA11p12023}

\bibitem[{{Roberts} {et~al.}(1992){Roberts}, {Goldstein}, {Matthaeus}, \&
  {Ghosh}}]{RobertsEA1992}
{Roberts}, D.~A., {Goldstein}, M.~L., {Matthaeus}, W.~H., \& {Ghosh}, S. 1992,
  \jgr, 97, 17115, \dodoi{10.1029/92JA01144}

\bibitem[{{Ruffolo} {et~al.}(2021){Ruffolo}, {Ngampoopun}, {Bhora},
  {Thepthong}, {Pongkitiwanichakul}, {Matthaeus}, \& {Chhiber}}]{Ruffolo2021}
{Ruffolo}, D., {Ngampoopun}, N., {Bhora}, Y.~R., {et~al.} 2021, \apj, 923, 158,
  \dodoi{10.3847/1538-4357/ac2ee3}

\bibitem[{{Russell}(1972)}]{Russell72}
{Russell}, C.~T. 1972, in NASA Special Publication, ed. C.~P. {Sonett}, P.~J.
  {Coleman}, \& J.~M. {Wilcox}, Vol. 308, 365

\bibitem[{{Sanchez-Diaz} {et~al.}(2019){Sanchez-Diaz}, {Rouillard}, {Lavraud},
  {Kilpua}, \& {Davies}}]{Sanchez-DiazEA19}
{Sanchez-Diaz}, E., {Rouillard}, A.~P., {Lavraud}, B., {Kilpua}, E., \&
  {Davies}, J.~A. 2019, \apj, 882, 51, \dodoi{10.3847/1538-4357/ab341c}

\bibitem[{{Servidio} {et~al.}(2008){Servidio}, {Matthaeus}, \&
  {Dmitruk}}]{ServidioEA08}
{Servidio}, S., {Matthaeus}, W.~H., \& {Dmitruk}, P. 2008, \prl, 100, 095005,
  \dodoi{10.1103/PhysRevLett.100.095005}

\bibitem[{{Shebalin} {et~al.}(1983){Shebalin}, {Matthaeus}, \&
  {Montgomery}}]{ShebalinEA83}
{Shebalin}, J.~V., {Matthaeus}, W.~H., \& {Montgomery}, D. 1983, Journal of
  Plasma Physics, 29, 525, \dodoi{10.1017/S0022377800000933}

\bibitem[{{Shi} {et~al.}(2023){Shi}, {Sioulas}, {Huang}, {Velli}, {Tenerani},
  \& {R{\'e}ville}}]{Shi_2023}
{Shi}, C., {Sioulas}, N., {Huang}, Z., {et~al.} 2023, arXiv e-prints,
  arXiv:2308.12376, \dodoi{10.48550/arXiv.2308.12376}

\bibitem[{Shi {et~al.}(2021)Shi, Velli, Panasenco, Tenerani, Réville, Bale,
  Kasper, Korreck, Bonnell, Dudok~de Wit, Malaspina, Goetz, Harvey, MacDowall,
  Pulupa, Case, Larson, Verniero, Livi, Stevens, Whittlesey, Maksimovic, \&
  Moncuquet}]{Shi_2021}
Shi, C., Velli, M., Panasenco, O., {et~al.} 2021, Astronomy and astrophysics,
  650, A21, \dodoi{10.1051/0004-6361/202039818}

\bibitem[{{Smith}(2001)}]{Smith2001}
{Smith}, E.~J. 2001, \jgr, 106, 15819, \dodoi{10.1029/2000JA000120}

\bibitem[{{Stribling} \& {Matthaeus}(1991)}]{StriblingMatthaeus91}
{Stribling}, T., \& {Matthaeus}, W.~H. 1991, Physics of Fluids B, 3, 1848,
  \dodoi{10.1063/1.859654}

\bibitem[{{Taylor}(1938)}]{Taylor38}
{Taylor}, G.~I. 1938, Proceedings of the Royal Society of London Series A, 164,
  476, \dodoi{10.1098/rspa.1938.0032}

\bibitem[{{Ting} {et~al.}(1986){Ting}, {Montgomery}, \& {Matthaeus}}]{TingEA86}
{Ting}, A.~C., {Montgomery}, D., \& {Matthaeus}, W.~H. 1986, Physics of Fluids,
  29, 3261, \dodoi{10.1063/1.865843}

\bibitem[{Tu \& Marsch(1993)}]{Tu_Marsch_1993}
Tu, C.-Y., \& Marsch, E. 1993, Journal of geophysical research, 98,
  1257–1276, \dodoi{10.1029/92ja01947}

\bibitem[{Walén(1944)}]{Walen_1944}
Walén, C. 1944, Arkiv for Matematik, Astronomi och Fysik, 30A, 1–87.
\newblock \url{https://ui.adsabs.harvard.edu/abs/1944ArMAF..30A...1W/abstract?}

\bibitem[{{Wang} {et~al.}(2011){Wang}, {Boldyrev}, \& {Perez}}]{WangEA11}
{Wang}, Y., {Boldyrev}, S., \& {Perez}, J.~C. 2011, \apjl, 740, L36,
  \dodoi{10.1088/2041-8205/740/2/L36}

\bibitem[{Whittlesey {et~al.}(2020)Whittlesey, Larson, Kasper, Halekas,
  Abatcha, Abiad, Berthomier, Case, Chen, Curtis, Dalton, Klein, Korreck, Livi,
  Ludlam, Marckwordt, Rahmati, Robinson, Slagle, Stevens, Tiu, \&
  Verniero}]{Whittlesey_2020}
Whittlesey, P.~L., Larson, D.~E., Kasper, J.~C., {et~al.} 2020, The
  Astrophysical journal. Supplement series, 246, 74,
  \dodoi{10.3847/1538-4365/ab7370}

\bibitem[{{Wu} {et~al.}(2021){Wu}, {Tu}, {Wang}, {He}, {Yang}, \&
  {Yao}}]{WuEA21}
{Wu}, H., {Tu}, C., {Wang}, X., {et~al.} 2021, \apj, 912, 84,
  \dodoi{10.3847/1538-4357/abf099}

\bibitem[{{Zank} {et~al.}(1996){Zank}, {Matthaeus}, \& {Smith}}]{ZankEA96}
{Zank}, G.~P., {Matthaeus}, W.~H., \& {Smith}, C.~W. 1996, \jgr, 101, 17093,
  \dodoi{10.1029/96JA01275}

\bibitem[{Zhao {et~al.}(2022)Zhao, Zank, Adhikari, Telloni, Stevens, Kasper,
  Bale, \& Raouafi}]{Zhao_2022}
Zhao, L.-L., Zank, G.~P., Adhikari, L., {et~al.} 2022, The astrophysical
  journal. Letters, 934, L36, \dodoi{10.3847/2041-8213/ac8353}

\bibitem[{{Zhou} \& {Matthaeus}(1989)}]{ZhouMatt89}
{Zhou}, Y., \& {Matthaeus}, W.~H. 1989, \grl, 16, 755,
  \dodoi{10.1029/GL016i007p00755}

\end{thebibliography}
\bibliographystyle{aasjournal}



\end{document}